\documentclass[prd, aps, superscriptaddress, tightenlines, nofootinbib, eqsecnum, amsfonts, amsmath, amssymb, floatfix, notitlepage,twocolumn]{revtex4-1}  
\pdfoutput=1

\usepackage[utf8]{inputenc} 
\usepackage{euscript}
\usepackage{epsfig}
\usepackage{graphics}
\usepackage{graphicx}
\usepackage{amsmath}
\usepackage{amssymb}
\usepackage{bm}
\usepackage[usenames,dvipsnames,svgnames,table]{xcolor}
\usepackage{xspace}
\usepackage{times}
\usepackage{appendix}
\usepackage{lipsum}
\usepackage[nolist,nohyperlinks]{acronym}
\usepackage{float}
\usepackage{simplewick}
\usepackage{tabularx}
\usepackage{booktabs}
\usepackage[normalem]{ulem}
\usepackage{natbib, ifthen}
\usepackage[hidelinks]{hyperref}
\usepackage{comment}
\usepackage{dsfont}
\usepackage[caption=false]{subfig}
\usepackage{soul}
\usepackage{bbold}
\usepackage{xfrac}
\usepackage{tikz}       

\newcommand{\fett}[1]{\boldsymbol{#1}}
\newcommand{\dd}{{\rm{d}}}

\newcommand{\be}{\begin{equation}}
\newcommand{\ee}{\end{equation}}

\newcommand{\nabq}{\boldsymbol{\nabla}_{\!q}}
\newcommand{\nabr}{\fett{\nabla}_{\!r}}

\newcommand{\sdfrac}[2]{\mbox{\small$\displaystyle\frac{#1}{#2}$}}

\newcommand{\rRGone}{r_{\text{\fontsize{6}{6}\selectfont \!$1$RG}}}
\newcommand{\rRGtwo}{r_{\text{\fontsize{6}{6}\selectfont \!$2$RG}}}
\newcommand{\rRGfour}{r_{\text{\fontsize{6}{6}\selectfont \!$4$RG}}}
\newcommand{\RG}[1]{{\text{\fontsize{6}{6}\selectfont \!${#1}$RG}}}

\newcommand{\psiASY}{\psi_{\text{\fontsize{6}{6}\selectfont $\infty$}}}
\newcommand{\psiNLPTplusUV}[1]{\psi_{\text{\fontsize{6}{6}\selectfont $#1$UV}}}
\newcommand{\psiUVcounter}[1]{\psi_{\text{\fontsize{6}{6}\selectfont $\infty$}}^{(\!#1\!)}}
\newcommand{\NL}{\text{\fontsize{6}{6}\selectfont NL}}
\newcommand{\deltaNL}[1]{\delta_{\text{\fontsize{6}{6}\selectfont #1}}}
\newcommand{\deltalin}{\delta_{\rm lin}}

\definecolor{darkgreen}{rgb}{0,0.5,0}

\newcommand{\mnras}{Mon. Not. Roy. Astron. Soc.}
\newcommand{\apjs} {Astrophys. J. Suppl.}

\newcommand{\physrep} {Phys. Rep.}
\newcommand{\aap}{Astron. Astrophys.}

\newcommand{\jcap}{J. Cosmol. Astropart. Phys.}

\definecolor{lime}{HTML}{A6CE39}
\DeclareRobustCommand{\orcidicon}{
	\begin{tikzpicture}
	\draw[lime, fill=lime] (0,0) 
	circle [radius=0.14] 
	node[white] {{\fontfamily{qag}\selectfont \tiny ID}};
	\draw[white, fill=white] (-0.0625,0.095) 
	circle [radius=0.007];
	\end{tikzpicture}
	\hspace{-2mm}
}

\foreach \x in {A, ..., Z}{%
	\expandafter\xdef\csname orcid\x\endcsname{\noexpand\href{https://orcid.org/\csname orcidauthor\x\endcsname}{\noexpand\orcidicon}}
}

\begin{document}

\title[RG and UV completion]{Renormalization group and UV completion of cosmological perturbations: \\ 
 Gravitational collapse as a critical phenomenon}

\author{Cornelius Rampf${\,{\tiny\orcidA{}}}$\phantom{II}}\email{cornelius.rampf@univie.ac.at}
\affiliation{Department of Astrophysics, University of Vienna, T\"urkenschanzstraße 17, 1180 Vienna, Austria}
\affiliation{Department of Mathematics, University of Vienna, Oskar-Morgenstern-Platz 1, 1090 Vienna, Austria}

\author{Oliver Hahn${\,{\tiny\orcidC{}}}$\phantom{II}}\email{oliver.hahn@univie.ac.at}
\affiliation{Department of Astrophysics, University of Vienna, T\"urkenschanzstraße 17, 1180 Vienna, Austria}
\affiliation{Department of Mathematics, University of Vienna, Oskar-Morgenstern-Platz 1, 1090 Vienna, Austria}

\date{\today}

\begin{abstract}
Cosmological perturbation theory is known to converge poorly for predicting the spherical collapse and void evolution of collisionless matter. Using the exact parametric solution as a testing ground, we develop two asymptotic methods in spherical symmetry that resolve the gravitational evolution to much higher accuracy than Lagrangian perturbation theory (LPT), which is the current gold standard in the literature. One of the methods selects a stable fixed-point solution of the renormalization-group flow equation, thereby predicting already at the leading order the critical exponent of the phase transition to collapsed structures. The other method completes the truncated LPT series far into the UV regime,  by adding a non-analytic term  that captures the critical nature of the gravitational collapse. We find that the UV method most accurately resolves the evolution of the nonlinear density as well as its one-point probability distribution function. Similarly accurate predictions are achieved with the renormalization-group method, especially when paired with Pad\'e approximants. Further, our results yield new, very accurate, formulas to relate linear and nonlinear density contrasts. Finally, we chart possible ways on how to adapt our methods to the case of cosmological random field initial conditions.
\end{abstract}


\maketitle

\section{Introduction}

Current and upcoming surveys of the cosmic large-scale structure  are expected to test the cosmological concordance model at a precision of about one percent \cite{2011MNRAS.416.3017B,2016MNRAS.460.1270D,2018PASJ...70S...8A,2009arXiv0912.0201L,2011arXiv1110.3193L}. 
Therefore, having fast and accurate theoretical predictions for the large-scale structure is of fundamental importance.
For example, this is relevant in the context of field-level forward models 
for reconstruction of tracers of the matter distributions, such as retrieved from
the galaxy density field \cite{2013MNRAS.432..894J,2013MNRAS.429L..84K,2013ApJ...772...63W,2021MNRAS.500.3194A}
or from quasar spectra obtained from the Lyman-$\alpha$ forest data \cite{2012MNRAS.420...61K,2018MNRAS.477.2841M,2020A&A...642A.139P,2020JCAP...07..010R}.
Another important application relates to pairing theoretical predictions with numerical simulations \cite{2013JCAP...06..036T,2015A&C....12..109H,2016MNRAS.463.2273F,2021MNRAS.503.1897C,2021MNRAS.505.1422K,2021arXiv210112187Z,2021arXiv210414568A}. 

Cosmological perturbation theory (CPT; \cite{1980lssu.book.....P,1984ApJ...279..499F,Bernardeau2002,2014JCAP...01..010B,2021RvMPP...5...10R}) provides highly accurate predictions on large cosmological scales, and in particular plays a crucial role in connecting early- with late-time cosmology, such as by providing initial conditions for numerical simulations \cite{1983MNRAS.204..891K,1985ApJS...57..241E,1998MNRAS.299.1097S,2006MNRAS.373..369C,2021MNRAS.500..663M}. However, CPT struggles to accurately predict the small-scale collapse to gravitationally bound structures, a process that is intimately tied to the shell-crossing singularity---the crossing of trajectories of collisionless matter, which 
comes with extreme matter densities.

To remedy the problem,  many different approaches beyond standard CPT have been developed, for example by applying several renormalization or resummation techniques, as well as effective field-theory methods or path-integral formalisms (see e.g.\ \cite{2004A&A...421...23V,2006PhRvD..73f3519C,2007PhRvD..75d3514M,2007JCAP...06..026M,2008MPLA...23...25M,2008JCAP...10..036P,2008PhRvD..77f3530M,2012JHEP...09..082C,2013MNRAS.429.1674C,2013JCAP...09..024B,2019AnP...53100446B}). However, these approaches are either of statistical nature and thus cannot be directly tested against deterministic solutions or, to our knowledge, such critical tests have not yet been performed (see however \cite{2016JCAP...01..043M} for an exception).
Other approaches circumvent the shortcomings of CPT, by combining CPT predictions with those of the spherical or ellipsoidal collapse model \cite{Kitaura:2013,Bernardeau:1994,Mohayaee:2006,Monaco:2002,Monaco:2013,Stein:2019,Neyrinck:2016,Tosone:2021} by performing a large-scale/small-scale split---see e.g.\ Ref.\,\cite{Monaco:2016} for a review, and Ref.\,\cite{Lippich:2019} for performance tests of such hybrid methods for modeling galaxy clustering.

Still, these studies do not attempt to tackle the obvious problem, namely that CPT is highly inefficient for predicting the nonlinear collapse.
We believe that this problem has been partially addressed by  Refs.\,\cite{2007PhRvD..75d4028T,1998ApJ...498...48Y,1998ApJ...504....7M,2022PhRvD.105j3507T}, who tested Pad\'e approximants or Shanks transforms to accelerate the convergence of CPT.
As we will outline in this article, even a faster convergence acceleration can be achieved, provided one exploits the asymptotic structure of the underlying collapse problem.

In all these matters, the choice of coordinates can be crucial.
Indeed, the first nontrivial shell-crossing solutions have been found using Lagrangian coordinates \cite{2017MNRAS.471..671R,2018PhRvL.121x1302S,2022A&A...664A...3S}, specifically within the framework of Lagrangian perturbation theory (LPT). 
Reasons for LPT performing better in comparison to its Eulerian counterpart are various, here we just name two: First, Lagrangian approaches are very efficient in resolving advected motion, essentially since the involved nonlinearities are absorbed into the Lagrangian time derivative. Second, the Lagrangian map acts as a de-singularization transformation, meaning that the infinity in the matter density at shell crossing gets converted into a vanishing of the Jacobian determinant. Obviously, resolving the latter is technically easier than the former, especially within the context of perturbation theory.

Despite these advantages, however, even LPT converges extremely slowly for spherical or quasispherical collapse \cite{1994ApJ...436..517M,1998ApJ...498...48Y,2019MNRAS.484.5223R,2018PhRvL.121x1302S,2022A&A...664A...3S}.
Another, somewhat surprising, limitation of LPT is related to the late-time evolution of voids, which shows clear signs of loss of convergence (e.g.\ \cite{1996MNRAS.282..641S,1997A&A...326..873K}). This divergent behavior has been analyzed by the concept of  ``mirror symmetry'' \cite{Nadkarni-Ghosh:2011,Nadkarni-Ghosh:2013}, 
which elucidates that the radius of convergence of the LPT series is identical for both over- and underdense regions, essentially since the radius of convergence is independent of the sign of the spatial curvature parameter.

In this article, we revisit in detail the convergence issues of LPT for spherical over- and underdensities, and provide a systematic analysis of two asymptotic techniques. We will see that the poor convergence of the LPT series can be remedied by employing a technique dubbed UV completion, and this twofold:
First, the convergence of the UV-completed LPT series is vastly accelerated in overdense regions and, second, the underdense evolution does not display any divergent behavior anymore. At the core, the UV completion exploits the asymptotic structure of the LPT series at order infinity. The functional form of this asymptotic behavior appears to be quite generic \cite{2021MNRAS.501L..71R}, which raises the justified hope that this technique might be  adaptable in the near future also to the case of gravitational collapse with cosmological random field initial conditions.

In addition to the UV completion, we also study an alternative approach by applying a renormalization group (RG) technique to the spherical-collapse problem.
Historically, the RG approach was introduced in quantum electrodynamics to regularize infinities in quantum field theory. 
RG is also vital in statistical physics, for example when investigating critical phenomena in phase transitions  \cite{Wilson:1971dc}, which is also an instance where singularities appear (e.g., in the heat capacity).
RG techniques are also heavily employed in non-equilibrium physics to detect singularities  (e.g.\ \cite{Goldenfeld:1992qy,1994PhRvL..73.1311C,Chen:1996,2008PhyD..237.1029D,2008PhRvE..77a1105K}), which is particularly relevant for investigating critical phenomena in general fluids.
Crucially, as we show here by means of the spherical-collapse problem, the mere knowledge of the underlying singular structure provides a superior theoretical model as obtained through conventional perturbative techniques.

This paper is organized as follows. In the following we provide the basic fluid equations, first for generic initial conditions (Sec.\,\ref{sec:fluideqs}) and then limited to spherical symmetry (Sec.\,\ref{sec:sphericalsymEQS}). Afterwards, in Sec.\,\ref{sec:LPT}, we briefly review LPT.
In~Sec.\,\ref{sec:RG}, we discuss a particularly simple implementation of a renormalization-group approach (see App.\,\ref{app:RG+} for complementary derivations exploiting the RG-flow condition), including also the pairing of RG and Pad\'e approximants (Sec.\,\ref{sec:HotRG-Pade}).
Then, we provide an asymptotic analysis of the LPT series (Sec.\,\ref{sec:AsymLPT}), whose findings are then implemented to achieve a UV completion of the LPT series (Sec.\,\ref{sec:UVcompletion}).
Subsequently, we apply our findings to the calculation of the nonlinear density (Sec.\,\ref{sec:NLdensity}) and of its one-point probability distribution function (Sec.\,\ref{sec:PDF}). Finally, we summarize our findings and provide concluding remarks in Sec.\,\ref{sec:concl}.

\section{Basic setup}

\subsection{Fluid equations for generic initial data}\label{sec:fluideqs}

For simplicity, throughout this work,  we ignore the effects of a cosmological constant, and  focus solely on the nonlinear evolution of  collisionless matter.  
We employ Lagrangian coordinates $\fett q$ that denote the positions of fluid elements at initial time~$t= t_0$. Likewise,
$\fett r \!=\! \fett r (\fett q,t)\! =\! \fett x(\fett q, t)/ a(t)$  is the physical position of a given fluid element 
at current time~$t$, while~$a$ is the cosmic scale factor normalized to unity at time~$t_0$. Using these definitions, the equations of motion of the fluid elements can be written as
\be \label{eqs:Laginter}
   \ddot {\fett{r}} =  -\nabr \phi ,  \qquad \nabr^2 \phi = 4\pi G \rho , \qquad  \rho = \bar \rho / J ,
\ee
where an overdot stands for a Lagrangian (convective) time derivative, while $\bar \rho = \bar \rho_0 a^{-3}$ is the spatially uniform background density with $\bar \rho_0 = \bar \rho (t_0)$ denoting its initial value.
Furthermore, we have defined 
$J := \det[x_{i,j}]$ where a ``$,j$'' is a partial space derivative with respecect to the Lagrangian component~$q_j$.

Equations~\eqref{eqs:Laginter} can be easily merged into a single one by taking the Eulerian divergence from the first of the equations; see e.g.~Ref.\,\cite{2012JCAP...06..021R} for calculational details. After converting the remaining Eulerian derivative according to $\nabr = [(\nabq \fett r)^*]^{-1} \nabq$ where the star denotes matrix transposition, one arrives at the main evolution equation in Lagrangian coordinates \cite{1987JMP....28.2714B,1989A&A...223....9B}
\be \label{eq:mainLag}
  \varepsilon_{ilm} \varepsilon_{jpq} \,r_{p,l} \, r_{q,m} \, \ddot r_{j,i} = -8\pi G \bar \rho_0 \,.
\ee
Here, summation over repeated indices is assumed, and  $\varepsilon_{ilm}$ is the fundamental antisymmetric tensor. 
This scalar equation should be supplemented with the statement of the conservation of the vanishing vorticity (see, e.g., \cite{1994MNRAS.267..811B,2015MNRAS.452.1421R}), which however is not needed in what follows, due to the assumed symmetry.

\subsection{The case of spherical symmetry}\label{sec:sphericalsymEQS}

Equation~\eqref{eq:mainLag} is valid for any initial data, but in the following we will focus exclusively on the case of spherical symmetry. 
For this, one may employ spherical coordinates, but we find it actually more convenient to stick with the  Cartesian setup, in which
case the Jacobian matrix  $\nabq \fett r$ must be exactly diagonal with identical entries (e.g.\ \cite{2005pfc..book.....M}). Thus, for the components of the Jacobian matrix,  we can set 
\be
  r_{i,j} =\delta_{ij} \, r \,,
\ee
where $r$ depends only on time but not on space, and $\delta_{ij}$ is the Kronecker delta. With this simplification, Eq.\,\eqref{eq:mainLag} becomes
\be \label{eq:EmdenFowler}
  \boxed{ r^2  \, \ddot r  = - \frac{4\pi G \bar \rho_0}{3} } \,.
\ee
This is an ordinary differential equation of the Emden--Fowler type~\cite{1995heso.book.....P}, for which an exact parametric solution is well known (e.g.\ \cite{1934rtc..book.....T,1967ApJ...147..859P,1969PThPh..42....9T,1972ApJ...176....1G,1994ApJ...431..486B}); see App.\,\ref{app:para} for a review as well as asymptotic considerations. Having access to an exact solution provides an ideal testing ground for developing new solutions techniques. In the following, we first briefly review LPT with its shortcomings in the context of spherical collapse.

\section{Lagrangian perturbation theory}\label{sec:LPT}

The central quantity in LPT is the displacement field 
\be
  \fett \psi(\fett q,t) =\fett{x}(\fett q, t) - \fett q\,,
\ee 
which in the case of spherical symmetry can be written in a simplified manner as $\fett \psi(\fett q,t) = \fett q\,\psi(t)$, where the scalar displacement $\psi(t)$ depends only on time.
The fastest-growing mode of this displacement is expanded as a perturbative series (e.g.~\cite{1991ApJ...382..377M,1993MNRAS.264..375B,1995A&A...296..575B,1997GReGr..29..733E,2014JFM...749..404Z,2018PhRvL.121x1302S,2019MNRAS.484.5223R}), 
\be \label{eq:x}
 \psi(t) =  \sum_{n=1}^\infty \psi_n (k a)^n \,,
\ee
where $\psi(t) = x(t)-1$, and the (scalar) comoving trajectory $x(t)$ is defined via $x_{i,j} = \delta_{ij}x$.
Furthermore,  $k$ is a free parameter fixed by the initial conditions which, physically, amounts to an effective curvature parametrizing the local departure from a spatially flat Universe; see e.g.\ Ref.\,\cite{2019MNRAS.484.5223R} for calculational details.

In LPT, the {\it Ansatz}~\eqref{eq:x} is used to solve Eq.\,\eqref{eq:EmdenFowler} at subsequent perturbation orders $n$.
The $\psi_n$'s occurring in Eq.\,\eqref{eq:x} are easily determined by the following recursive relations ($n \geq 1$) \cite{2019MNRAS.484.5223R}
\begin{align} \label{eq:rec}
 \psi_n &= -\sdfrac{1}{3} \delta_{n1}  
   - \sum_{q<n}  \sdfrac{q^2 + (n-q)^2 - (3-n)/2  }{(n + 3/2) ( n-1)} \psi_q \psi_{n-q} \nonumber \\
   & -  \sum_{k+l+m=n} \sdfrac{ k^2 + l^2 + m^2 - (3-n)/2}{3(n + 3/2) ( n-1)}  
   \psi_k \psi_l \psi_m \,,
\end{align}
which in particular leads to the first few coefficients
\begin{align} \label{eq:psis}
 \psi_1 &= - \sdfrac 1 3  \,, \qquad  \psi_2 =  -\sdfrac{1}{21}\,, \qquad  \hspace{0.1cm}
  \psi_3 = -\sdfrac{23}{1701} \,. 
\end{align}
See Sec.\,\ref{sec:asy-UV} for investigating the large-order asymptotic properties of the displacement series.

\begin{figure}
 \centering
   \includegraphics[width=0.99\columnwidth]{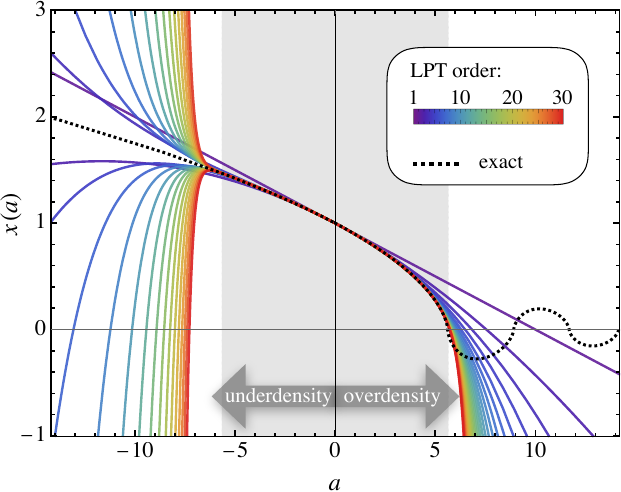}
   \caption{LPT series solutions for the comoving trajectory $x(a) = 1 + \psi(a)$ up to 30th order for the case $k\!=\!3/10$ (colored lines), compared against the exact solution (black dotted line) based on the spherical collapse model, which begins oscillating around $x=0$ after the first collapse. The positive $a$-branch denotes the collapse of a spherical overdensity, while the negative $a$-branch reflects the void evolution with $k=-3/10$. The  gray-shaded area indicates the disk of convergence, i.e., the region $-a_\star<a<a_\star$ where $a_\star=(3\pi\sqrt{2})^{2/3} \simeq 5.622$ is the collapse time of the overdensity. 
Clearly, convergence of the LPT series is lost for $|a|> a_\star$ for both over- and underdense regions.
}   \label{fig:rainbow}
\end{figure}

In Fig.\,\ref{fig:rainbow} we show the comoving trajectory $x(a) = 1+ \psi(a)$ for several LPT truncation orders (various colored lines), and compare it against the exact parametric solution (black dotted line, based on App.\,\ref{app:para}). Specifically, the positive $a$-time branch depicts the collapse of a spherically overdense region with curvature $k=3/10$, while the negative time branch reflects the  evolution of an underdense region with $k\!=\!-3/10$ where, for convenience, we exploit the sign symmetry of the tuple $(a,-k) \leftrightarrow (-a,k)$ within the definition of $\psi(a)$. 
In the present case, LPT convergence is lost at the critical time $|a|=a_\star =(3\pi\sqrt{2})^{2/3} \simeq 5.622$, 
which coincides exactly with the time of spherical collapse. More generally, for given curvature $k$, the time of collapse is $a_\star = \delta_{\rm c}/k$, where $\delta_{\rm c} = (3/5) (3\pi/2)^{2/3} \simeq 1.686$ is the usual critical density at collapse time (see e.g.\ \cite{1967ApJ...147..859P,1969PThPh..42....9T,1972ApJ...176....1G,1994ApJ...431..486B}).

Moreover, since LPT is a Taylor series in powers of $a$, the temporal regime of convergence is spanned by $|a| < a_\star$, 
where the radius of convergence  $a_\star$ is set by the nearest singularity(ies) in the complex-time plane around the expansion point $a=0$. 
In the case of spherical symmetry, the nearest singularity occurs at the real-valued  collapse time $a_\star$ where, specifically, the velocity $v= \dot x$ blows up \cite{2019MNRAS.484.5223R,2021RvMPP...5...10R}.
This explains the exact coincidence between the collapse time and the loss of convergence in the case of spherical symmetry.

The above argument also implies that the LPT series in the void case must have an identical radius of convergence of~$a_\star$, which is also shown in Fig.\,\ref{fig:rainbow} in the negative time branch (exploiting the aforementioned sign symmetry). Clearly, the lack of convergence in the void case beyond $a_\star$ is a mathematical artifact since, physically, the void underdensity should not be influenced by a singular velocity appearing in the spherical collapse. To our knowledge, the fact that the LPT series for over- and underdensities have an identical radius of convergence was first pointed out in Ref.\,\cite{Nadkarni-Ghosh:2011}.

\section{Renormalization-group approach} \label{sec:RG}

Multiplying Eq.\,\eqref{eq:EmdenFowler} by $\dot r/r^2$ and integrating the resulting equation in time, one obtains  
\be
  \dot r^2 = \frac{8\pi G \bar\rho_0}{3r}  - C,
\ee
where $C$ is an integration constant.
Now, changing from cosmic time to scale-factor time $a$ with $\partial_t = \dot a \partial_a$ and using the
Friedmann equation $(\dot a/a)^2 = 8\pi G \bar \rho_0/(3a^3)$ to get an expression for~$\dot a$,
 we arrive after some straightforward calculations 
at our main evolution equation in the RG approach,
\be \label{eq:RGmain}
  \boxed{ r'^2 = \frac a r - \epsilon \frac a 2 } \,,
\ee
where a prime denotes a partial derivative w.r.t.\ $a$-time, and we have defined $\epsilon = 3C/(4\pi G \bar \rho_0)$. Actually, in the RG approach $\epsilon$ acts as a perturbative bookkeeping parameter; physically, it can be interpreted as a curvature perturbation locally induced through a spherical over- or underdense region in an otherwise spatially flat Universe (see also App.\,\ref{app:para}). 
In the following, we seek solutions for~\eqref{eq:RGmain} by means of a standard perturbative expansion, i.e., 
\be \label{eq:RGans}
  r = r_0 + \epsilon \, r_1 + \epsilon^2 \, r_2 + \ldots \,.
\ee
This perturbative {\it Ansatz} allows us to obtain a set of differential equations for $r_0$, $r_1$, etc.\ that are easily integrated in time.
The resulting approximation for $r$ will have so-called secular terms that grow unboundedly for large times, which is an indication of ill-posed asymptotic behavior. 
To  cure the perturbative results from these secular terms, we then apply suitable renormalization techniques.

In the following we provide a reduced RG approach that, for the present problem, is particularly simple in its implementation; see App.\,\ref{app:RG+} for more traditional avenues exploiting the RG-flow equation, leading however to identical results as quoted here.

\subsection{Perturbative expansion and temporal integration}

Plugging the {\it Ansatz}~\eqref{eq:RGans} into the Eq.\,\eqref{eq:RGmain} and  keeping only terms to fixed orders in $\epsilon$, one
obtains the following set of differential equations:
\begin{align}
  &\epsilon^0 \bigg\{ r_0'^2 = \frac{a}{r_0}  \bigg\} \,, \label{eq:evoRG0} \\
  &\epsilon^1 \bigg\{ 2 r_0' r_1' + a \frac{r_1}{r_0^2} + \frac a 2 = 0 \bigg\}  \,, \label{eq:evoRG1}  \\
  &\epsilon^2 \bigg\{ 2 r_0' r_2' + r_1'^2 +  ( r_0 r_2 - r_1^2) \frac{a}{r_0^3} =0 \bigg\}  \,, \label{eq:evoRG2}
\end{align}
and so on, where we remind the reader that a prime denotes a temporal derivative w.r.t.\ $a$-time. 
Integrating the zeroth-order equation~\eqref{eq:evoRG0} leads to
\be \label{eq:R0}
   r_0 = a \left( 1 + \frac{3 c_1}{2 a^{3/2}} \right)^{2/3} \,,
\ee
where $c_1$ is an integration constant which plays a crucial role in the renormalization procedure exploited below (but otherwise would be determined by the initial conditions).
Continuing to the next orders, the first-order differential equations~\eqref{eq:evoRG1} and~\eqref{eq:evoRG2}
have particular solutions
\be
  r_1 =  - \frac {1}{10} r_0^2  \,,
 \qquad r_2 = -\frac{3}{700} r_0^3 \,,
\ee
respectively. Here, the homogeneous parts of the solutions can be discarded as they would 
add more integration constants than being allowed by the parent ODE~\eqref{eq:RGmain}. 
In any case,  higher-order homogeneous parts should not play any role in the physical solution.
See App.\,\ref{app:refinedRG} for a complementary RG method relying on specific boundary conditions.

\subsection{First-order renormalization}

Let us begin with a renormalization procedure that ignores the effects of $O(\epsilon^2)$.
In that case, the unrenormalized solution for $r$ reads
\be \label{eq:RG2}
  r  =  a \left( 1 + \frac{3 c_1}{2 a^{3/2}} \right)^{2/3} - \frac \epsilon {10} r_0^2 \,,
\ee
where the last term $r_1 \sim r_0^2$ is of secular nature.
To accommodate this term, 
we perform a multiplicative renormalization of the constant $c_1$, with the requirement that the shift should mimic the secular term $r_1$ to order $\epsilon$, i.e.,  we demand  $c_1 \to c_1 (1+ \epsilon A)$ and search for the unknown $A$. After straightforward calculations one finds
\be \label{eq:A}
  A = - \frac{1}{10 c_1}a^{5/2} \left( 1+ \frac{3 c_1}{2 a^{3/2}} \right)^{5/3} \,.
\ee
Plugging this back into~\eqref{eq:RG2} one finds 
\be
 r = a \left( 1 + \frac{3 c_1 \left[  1 - \frac{a^{5/2}}{10 c_1} \left(  1 + \frac{3 c_1}{2 a^{3/2}} \right)^{5/3} \epsilon  \right]  }{2 a^{3/2}}   \right)^{2/3}  \,,
\ee
to first order in $\epsilon$. Finally,
we discard decaying modes as they should have negligible impact in the asymptotic/late-time limit, and obtain
the first-order RG solution $r = \rRGone+ O(\epsilon^2)$, with
\be \label{eq:1RGsol}
  \rRGone = a \left(  1 - \frac {3 a \epsilon}{20} \right)^{2/3} \,.
\ee
In App.\,\ref{app:refinedRG} we show that the above derivation selects a stable fixed-point solution of the RG flow equation.

It is also interesting to note that a first-order Taylor expansion of $\rRGone$ about $a=0$ delivers 
 the first-order LPT result for $\epsilon=10k/3$. We will see that this is a generic property of 
the employed RG technique. However, we note that by Taylor-expanding the RG result to fixed order, one also loses its desirable asymptotic properties.

Indeed, the asymptotic behavior of the RG result is vastly different when compared to LPT at any fixed order: Specifically, RG predicts a singularity, i.e. non-analyticity (non-differentiability), at the collapse time of $a = a_{\star, \RG{1}} = 20/3 \simeq 6.67$ for $\epsilon =1$ [achieved by setting the round bracketed term in Eq.\,\eqref{eq:1RGsol} to zero]. 
This shell-crossing prediction, which is just a first-order approximation within the RG approach, outperforms the third-order LPT prediction ($a_{\star, \text{\fontsize{6}{6}\selectfont 3LPT}} \simeq 6.83$, as opposed to the exact result $a_\star \simeq 5.62$).
Furthermore and most interestingly, the 1RG solution comes with a critical exponent of~$2/3$, thereby predicting correctly the blowup of the velocity~$v = \dot x$ at shell-crossing time. We will comment on this further below (see also App.~\ref{app:para}).

\subsection{Second-order renormalization}

Collecting all terms up to $O(\epsilon^2)$, the  unrenormalized second order solution for $r$ is
\be
  \label{eq:RG3}
  r  =  a \left( 1 + \frac{3 c_1}{2 a^{3/2}} \right)^{2/3} - \frac \epsilon{10} r_0^2  - \frac{3 \epsilon^2}{700} r_0^3   \,.
\ee
Now we add a second-order renormalization term, i.e., $c_1 \to c_1 (1 + \epsilon A + \epsilon^2 B)$, with the task that the new unknown $B$  absorbs the secular terms at order $\epsilon^2$ in~\eqref{eq:RG3}.
Of course, the $A$ term arising from the first-order renormalization remains unchanged.
 We find
\be  \label{eq:B}
  B = -\frac{5}{2800 c_1} a^{7/2} \left( 1+ \frac{3c_1}{2a^{3/2}} \right)^{7/3} \,.
\ee
Plugging this into~\eqref{eq:RG3} one first obtains
\be
  r = a \left( 1 + \frac{3 c_1 \left[  1 + A \epsilon  + B \epsilon^2 \right]}{2 a^{3/2}}   \right)^{2/3}  
\ee
up to $O(\epsilon^3)$, 
where $A$ and $B$ are, respectively, given in Eqs.\,\eqref{eq:A} and~\eqref{eq:B}.
Discarding the decaying modes we then obtain the second-order RG solution $r= \rRGtwo + O(\epsilon^3)$, with
\be \label{eq:RG2sol}
  \rRGtwo = a \left(  1 - \frac {3 a \epsilon}{20}  -  \frac{3a^2 \epsilon^2 }{1120}\right)^{2/3}.
\ee 
Expanding this result about $a=0$ to second order, we recover the standard 2LPT result [Eq.\,\eqref{eq:x} including the first two terms in the sum] for $\epsilon=10k/3$, indicating
that the 2RG result aligns with the LPT results for sufficiently early times, as it should.

\begin{figure}
 \centering
   \includegraphics[width=0.99\columnwidth]{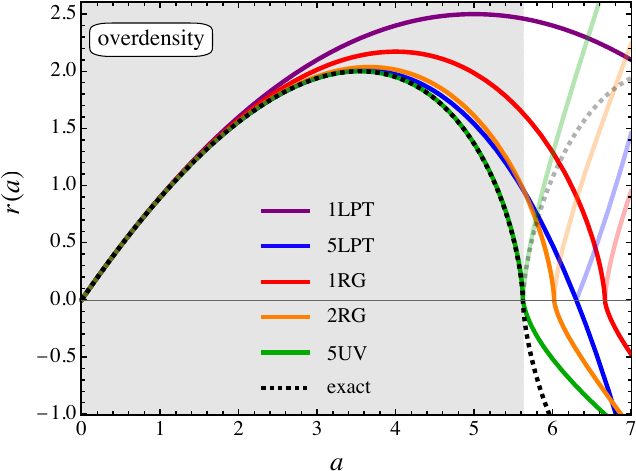}

   \vspace{0.5cm}

   \includegraphics[width=0.99\columnwidth]{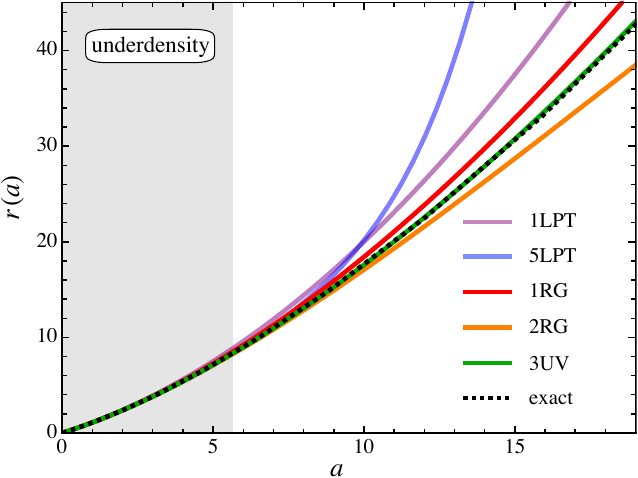}

   \caption{Temporal evolution of the physical trajectory $r(a)$ for $k=3/10$ (top panel) and for $k=-3/10$ (bottom panel), as predicted from various theoretical approaches. Specifically, ``1RG'' and ``2RG'' are based on Eqs.\,\eqref{eq:1RGsol} and~\eqref{eq:RG2sol} respectively for $\epsilon=\pm1$, while ``$n$UV'' is our novel UV approach discussed in Sec.\,\ref{sec:asy-UV}. The gray shading denotes the range of convergence of the LPT series~\eqref{eq:x}. Solid lines take the real values from the solutions, while fainter lines depict the continuations of the absolute value of the respective solutions.
}   \label{fig:over-underdense}
\end{figure}

In Fig.\,\ref{fig:over-underdense} we show the temporal evolution of the physical trajectory using the so-obtained 1RG and 2RG solutions (see Fig.\,\ref{fig:pade} for results up to 4RG). 
In the overdense case, the RG (and also the UV) solutions become complex after collapse, and therefore we show the real parts (solid lines/dots) as well as the absolute parts (faint lines/dots) of the respective solutions.
Clearly, 1RG and 2RG resolve the collapse to higher accuracy as compared to low-order LPT predictions (top panel).
In the underdense case (lower panel), 1RG and 2RG are performing well even for some time beyond the expected range of validity. Still, comparing the late-time evolution between the two RG solutions in the void case, it is observed that 2RG  does not improve significantly over 1RG for $a\gtrsim10$. 

In fact, going to times much later than shown in Fig.\,\ref{fig:over-underdense}, 2RG performs poorly. The reason for that is that the term $\sim a^2$ within the brackets of~\eqref{eq:RG2sol} adds a root in $\rRGtwo(a)$ at $a=a_\star = - \epsilon^{-1} (4/3) [21+ \sqrt{651}]\approx -62.02/\epsilon$, pushing the void solution into an artificial collapse and, therefore, into unphysical behavior.
We will further address related points in the following, as well as simple counter measures.

\subsection{Higher-order RG and Pad\'e approximants} \label{sec:HotRG-Pade}

It is straightforward to extend the formalism above to higher orders in $\epsilon$. For example, at order 4RG
we find
\begin{align} \label{eq:4RG}
  \rRGfour &= a {\cal X}^{2/3} 
\end{align}
to $O(\epsilon^5)$, where
\be \label{eq:calX}
 {\cal X}(a) := 1 - \frac {3 a \epsilon}{20}  -  \frac{3a^2 \epsilon^2}{1120} - \frac{11a^3 \epsilon^3}{67200} - \frac{823 a^4 \epsilon^4}{59136000} \,.
\ee
Generally, higher-order RG solutions become more accurate within the disk of convergence (gray shaded areas in Figs.\,\ref{fig:over-underdense}). However, as pointed out above, beyond the expected range of convergence---which is particularly relevant in the void case---higher-order RG exemplifies similar divergent behavior as observed in the LPT case.
Actually, such divergent behavior is  generally expected for asymptotic methods, i.e., such approaches do not necessarily perform better at higher iterations. There are two ways out of this dilemma, namely either to stay as low as possible in the perturbative iteration within the asymptotic method, or to investigate other means to remedy the shortcomings.

For this reason, we exploit in the following Pad\'e approximants to the interior term ${\cal X}$ appearing in the 4RG result. We remark that similar approximations have been already applied to the ``plain'' LPT series expansion \cite{1998ApJ...498...48Y,1998ApJ...504....7M,2007PhRvD..75d4028T}, but we stress that Pad\'e approximants  of the LPT series are vastly different to the ones that we apply to ${\cal X}$, 
essentially since in our RG approach the term ${\cal X}$ is exponentiated with $2/3$, thereby correctly capturing the leading-order asymptotic behavior of the LPT series. See also App.\,\ref{app:para} for further details and in particular Fig.\,\ref{fig:appVR}.

\begin{figure}
 \centering
   \includegraphics[width=0.99\columnwidth]{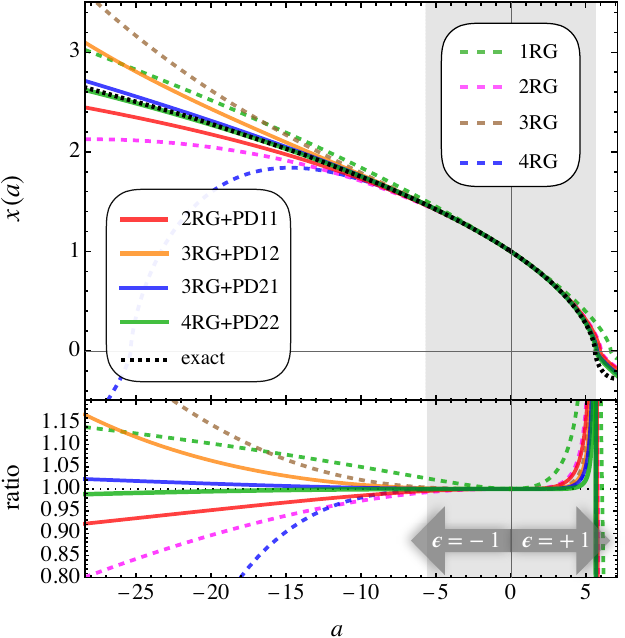}

   \caption{Temporal evolution of the comoving trajectory  $x = r/a$ for RG-related results for $\epsilon \!= \!\pm1$ (bottom panel: ratio of approximation versus exact result). 
  The long-dashed lines show pure RG results up to the fourth order, while the solid lines 
  involve additionally Pad\'e approximants. 
  Specifically, ``$i$RG+PD$mn$'' employs $r_{\RG{i}} /a= (P^{(m,n)} [ {\cal X}])^{2/3}$  for $i,m,n = 1,2,\ldots$, where the various 
  Pad\'e approximants are given in Eqs.\,\eqref{eqs:PDnm}.
  } \label{fig:pade}
\end{figure}

We define the Pad\'e approximant of degree $(m,n)$ for an arbitrary function $f(a)$ with
\be \label{eq:PDdef}
 P^{(m,n)}[\text{\small $f(a)$}] := \frac{\sum_{j=0}^m c_j a^j}{1+ \sum_{k=1}^n d_k a^k} \,,
\ee
where  $c_j$ and $d_k$ are time-independent coefficients that can be determined by Taylor expanding $f(a)$ about $a=0$. Doing so for ${\cal X}(a)$ as defined in~\eqref{eq:calX}, we find the following Pad\'e approximants
\begin{subequations} \label{eqs:PDnm}
\begin{align}
 \!\!\!\! P^{(1,1)}[\text{\small ${\cal X}$}] &=  \sdfrac{47}{5} + \sdfrac{2352}{5(a \epsilon - 56)} , \\
  \!\!\!\! P^{(1,2)}[\text{\small ${\cal X}$}] &=  \sdfrac{56 (1459 a \epsilon - 8460)}{a \epsilon (10640 + 327 a\epsilon)- 473760} , \\ 
  \!\!\!\! P^{(2,1)}[\text{\small ${\cal X}$}] &=  \sdfrac{a \epsilon (10640 - 327 a\epsilon) - 50400}{280(11a\epsilon - 180)} , \\
  \!\!\!\!  P^{(2,2)}[\text{\small ${\cal X}$}] &=  [120859200 + a\epsilon (1469453 a\epsilon - 29597400)]  \nonumber \\
&\hskip-1cm\times  \left[35(3453120+a\epsilon [2083 a\epsilon - 327672])\right]^{-1} . \label{eq:PD22}
\end{align}
\end{subequations}
In Fig.\,\ref{fig:pade} we show the comoving trajectory $x = r/a$ for pure RG results (dashed lines) as well as 
RG descriptions that also include Pad\'e approximants (solid lines).
While all pure RG methods predict well the overdense collapse with increasing accuracy for higher iterations, one can also observe  the above mentioned pathological prediction for the late-time evolution of voids.

In fact, one reason for the good performance of 4RG+PD22 is [and similarly for 3RG+PD21], that its two roots appear exclusively on the positive time axis, specifically at
$a=a_{\pm} 
\simeq 5.59/\epsilon, 14.45/\epsilon$ [at $a_{\pm}\simeq 5.75/\epsilon, 26.78/\epsilon$]. This is in stark contrast to 2RG, where a second root was added in the negative time branch, causing the void solution to undergo an unphysical collapse, thereby spoiling its long-term accuracy. Our results indicate that the use of Pad\'e approximants seems critical to maintain a certain symmetry---no roots at negative time---an aspect that deserves deeper mathematical investigation in future work.

\section{Asymptotic analysis and UV completion}\label{sec:asy-UV}

We have just seen above that RG techniques can be used to circumvent inherent limitations of the LPT series. 
Here we shall follow an orthogonal approach: Instead of avoiding LPT, we develop a framework that allows us to extract and exploit the asymptotic properties of the displacement series 
\be \label{eq:psiLPT}
  \psi(a) = \sum_{s=1}^\infty \psi_s \mathfrak{a}^s   \,, \qquad \mathfrak{a} := a k \,,
\ee
thereby providing a theoretical description that we call UV completion [see Eq.\,\eqref{eq:psis} for the first few coefficients~$\psi_s$].

The general idea of the UV completion is as follows [see Eq.\,\eqref{eq:psicompl} for the final result].
Suppose that there exists an explicit solution for the nonperturbative displacement field, valid deep in the ultraviolet (short wavelength) regime. The functional form of that nonperturbative displacement  might not be known in general, but let us assume that we know at least some of its ``intrinsic'' properties,
which are encapsulated in the singular term 
\be \label{eq:psi-asymptotic}
  \psiASY(a) \propto \left( \mathfrak{a}_\star - \mathfrak{a} \right)^\nu \,.
\ee
Here, $\nu$ is neither zero nor a positive integer, and $\mathfrak a_\star$ denotes a certain time value when a singularity in the problem arises. 
For example, if $\nu<0$, then the displacement itself would ``blow up'' at $\mathfrak{a}= \mathfrak{a}_\star$ which of course would be unphysical. If instead $\nu>0$ but $\nu\notin\mathbb{N}$,
then the displacement will remain bounded,
but the $n$th derivatives with $n>\lfloor \nu\rfloor$
exhibit singular behavior at $\mathfrak{a}= \mathfrak{a}_\star$, which is an instance of non-analyticity ($\psi\notin C^\infty$, i.e., the solution is not infinitely differentiable, and therefore not globally representable by a Taylor series).

Having the above in mind, we conjecture that the UV-completed solution of the LPT displacement field~\eqref{eq:psiLPT}  is then obtained by splitting off the non-analytic piece $\psi_\infty$ as follows,
\be  \label{eq:formUV}
 \boxed{  \psiNLPTplusUV{n}(a) = \sum_{s=1}^{n-1} \psi_s \mathfrak{a}^s  + \psiASY(a) - \psiUVcounter{n-1}(a)} \,.
\ee
Here, the first term on the r.h.s.\ is the truncated LPT series, while
 $\psiUVcounter{n-1}$ denotes the Taylor expansion of $\psiASY$ about $a=0$ to truncation order $n-1$. This last term is needed to avoid the double counting of certain low-order coefficients. We remark that such UV completing schemes are well known in the field of general fluid dynamics and asymptotic analysis;
early related work can be found e.g.\ in Refs.\,\cite{https://doi.org/10.1002/sapm195332183,1957RSPSA.240..214D,10.1093/qjmam/27.4.423}.

Before going into the calculations, let us briefly summarize the steps needed to determine the unknowns $\nu$ and $\mathfrak{a}_\star$. First of all, the physical meaning of these unknowns can be obtained by considering the Taylor expansion of~\eqref{eq:psi-asymptotic} about $a=0$, and comparing subsequent ratios of these Taylor coefficients with ratios of the LPT coefficients $\psi_n$. A straightforward analysis then reveals that $\mathfrak{a}_\star$ is nothing but the radius of convergence of the LPT series, if and only if the involved limit in d'Alembert's ratio test
\be \label{eq:ratiotest}
    \frac{1}{\mathfrak{a}_\star} = \lim_{n \to \infty} \frac{\psi_n}{\psi_{n-1}} 
\ee
exists. Similar  asymptotic considerations performed on the ``graphical'' level (Fig.\,\ref{fig:domb}) then lead to the conclusion that $\nu$ is related to the slope of the ratio of coefficients at order infinity.

\subsection{Leading-order asymptotics of LPT series}\label{sec:AsymLPT}

As outlined above, we first need to analyze the asymptotic properties of the LPT series before we can perform the UV completion.  To do so, we follow mostly the methodology of Refs.\,\cite{2019MNRAS.484.5223R,10.1093/qjmam/27.4.423}. See also~Refs.\,\cite{2021MNRAS.500..663M,2021MNRAS.501L..71R} where a similar analysis has been applied employing cosmological initial conditions.

Before proceeding  let us consider the following problem. While the LPT series~\eqref{eq:psiLPT} comprises an exact mathematical solution within its disc of convergence, in practice we can only generate a finite number of coefficients in the infinite series.
This raises the following question: given a finite number of (low-order) LPT series coefficients $\psi_n$, how can we determine its asymptotic properties, which are decided at order infinity?

To make progress we first consider the Taylor series representation of the singular term appearing in~\eqref{eq:psi-asymptotic}, which reads
\be
   \left( \mathfrak{a}_\star - \mathfrak{a}\right)^\nu = \mathfrak{a}_\star^\nu \sum_{n=0}^\infty c_n \mathfrak{a}^n \,, 
\ee
where we used a generalized binomial coefficient 
\be \label{eq:genBin}
   \qquad c_n  = \begin{pmatrix}  \nu \\ n \end{pmatrix} [-\mathfrak{a}_\star]^{-n} \,.
\ee
From the very definition of $c_n$, it is clear that the ratio of Taylor coefficients of the singular term is, for any $n > 2$, given exactly by~\cite{10.1093/qjmam/27.4.423} 
\be \label{eq:cnratio}
  \frac{c_n}{c_{n-1}} =  \frac{1}{\mathfrak{a}_\star} \left( 1 - (1+\nu) \frac 1 n \right)\,. 
\ee
It is important to observe that this ratio is linear in $1/n$. This linear relationship suggests 
 that the unknowns $\mathfrak{a}_\star$ and $\nu$ can be obtained by a graphical method, namely by drawing subsequent ratios of $c_n/c_{n-1}$ against $1/n$.

Now comes the crucial twist: if the above mentioned linear relationship persists also for the ratios of the  LPT coefficients $\psi_n$ for $n \to \infty$, then we can deduce that the large-$n$ asymptotic behavior is precisely described by $\psi_\infty$ as given in Eq.\,\eqref{eq:psi-asymptotic}. Even more,
 by drawing $\psi_n/\psi_{n-1}$  against $1/n$ and evaluating the $y$ intercept, one essentially applies the ratio test~\eqref{eq:ratiotest}. The described method traces back to the work of Domb and Sykes \cite{1957RSPSA.240..214D} in a fluid-mechanical context. 

\begin{figure}
 \centering
   \includegraphics[width=0.95\columnwidth]{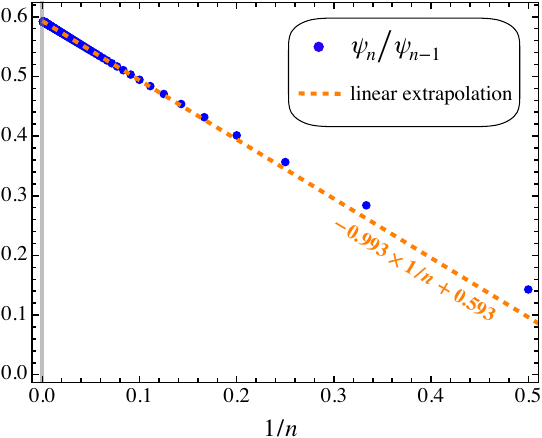}
   \caption{Domb--Sykes plot of subsequent ratios of coefficients $\psi_n/\psi_{n-1}$ over $1/n$ (blue data points) of the displacement series~\eqref{eq:psiLPT}. The coefficients $\psi_n$ are obtained through the recursive relation~\eqref{eq:rec}, and in the figure we show the results from the first 1000 coefficients.
The orange dashed line denotes a linear fit obtained between perturbation orders 900 and 1000. Extrapolation of this linear fit to the $y$ intercept (gray vertical line) reveals $\mathfrak{a}_\star \simeq 1/0.593 \simeq 1.686$, which coincides with the exact result for the shell-crossing time at the $10^{-6}$ level. 
}   \label{fig:domb}
\end{figure}

In Fig.\,\ref{fig:domb} we show the resulting ``Domb--Sykes'' plot for the LPT series~\eqref{eq:psiLPT}, obtained by drawing the subsequent ratios $\psi_{n}/\psi_{n-1}$ between the perturbation orders $n=2 - 1000$ (marked by blue dots). It is seen that these ratios settle into a linear relationship for sufficiently large orders, justifying a linear extrapolation to the $y$ intercept, from which one can read off the radius of convergence. In more detail, we apply a linear least-square fit between the perturbation orders $n=900 - 1000$, which reveals the linear model $-0.993 \times 1/n + 0.593$ (orange dashed line). Comparing this linear model against the form~\eqref{eq:cnratio}, one can read off the two ``free'' fitting parameters
\be \label{eq:DSresults}
  \mathfrak{a}_\star \simeq 1.686 \,, \qquad \nu \simeq 0.675 \,.
\ee
Here, $\mathfrak{a}_\star = a_\star k$ is identified as the radius of convergence, which coincides in the present case with the collapse time. From the spherical collapse model (Appendix~\ref{app:para}), we actually know the ``exact'' values for both parameters, namely $\mathfrak{a}_\star = \delta_{\rm c} = (3/5) [3\pi/2]^{2/3}$ and $\nu = 2/3$, where $\delta_{\rm c}$ is the linear density contrast at collapse time. While the above extrapolation technique is able to determine $\mathfrak{a}_\star$ at an accuracy of five significant digits, the ``error'' on determining the critical exponent is fairly large, namely about $1.22\,\%$. 
We note however that the numerical departure from $2/3$ could also arise due to the fact, that the extrapolation method detects corrections from the next-to-leading order asymptotic behavior (which originates from a new singular term with exponent $4/3$; see eq.\,\ref{eq:AsymSC}).
In any case,
as we shall see,
 even with the approximate value for $\nu$, the UV method 
delivers very accurate results (Figs.\,\ref{fig:over-underdense} and~\ref{fig:UV} employ the approximate value for $\nu$; see also App.\,\ref{app:UV} for further results).

\subsection{UV completion}\label{sec:UVcompletion}

The so-obtained results for $\mathfrak{a}_\star$ and $\nu$ from the asymptotic considerations can be directly used to extrapolate the LPT series to order infinity. To achieve this, we simply demand that the truncated LPT series to order $n$ is UV completed by the remainder of the series, where, crucially, the latter is proportional to the singular term $(\mathfrak{a}_\star - \mathfrak{a} )^\nu$. The amplitude of that singular term is naturally fixed by demanding that the series coefficients at the $n$th order are equal. This leads directly to 
\be \label{eq:psicompl}
  \boxed{ \psiNLPTplusUV{n}(a) = \sum_{s=1}^{n-1} \psi_s \mathfrak{a}^s + \frac{\psi_n}{c_n} \left[ \left(1-  \frac{\mathfrak{a}}{\mathfrak{a}_\star} \right)^\nu -  \sum_{k=0}^{n-1} c_k \mathfrak{a}^k \right] } \,,
\ee
where $\mathfrak{a} = a k$, and the generalized binomial coefficient $c_k$ is given in Eq.\,\eqref{eq:genBin}.
This is our final result, which can be used to determine the UV completion at a given truncation order~$n$.
For example, for $n=3$, we have 
\begin{align}
  \psiNLPTplusUV{3}(a) &= - \frac {\mathfrak{a}} 3  - \frac{\mathfrak{a}^2}{21}  + \frac{23\delta_{\rm c}}{756} \bigg\{  \mathfrak{a}^2 + 6 \mathfrak{a} \delta_{\rm c}   \nonumber \\
 &\qquad \hspace{1cm}+ 9 \left[  \left( 1 - \sdfrac{\mathfrak{a}}{\delta_{\rm c}} \right)^{2/3} \delta_{\rm c}^2 - 1 \right]  \bigg\} \,, \label{eq:3UV}
\end{align}
where  $\delta_{\rm c} = \mathfrak{a}_\star = (3/5) [3\pi/2]^{2/3} \simeq 1.686$, and, for simplicity, we set $\nu = 2/3$.

\begin{figure}
 \centering
   \includegraphics[width=0.99\columnwidth]{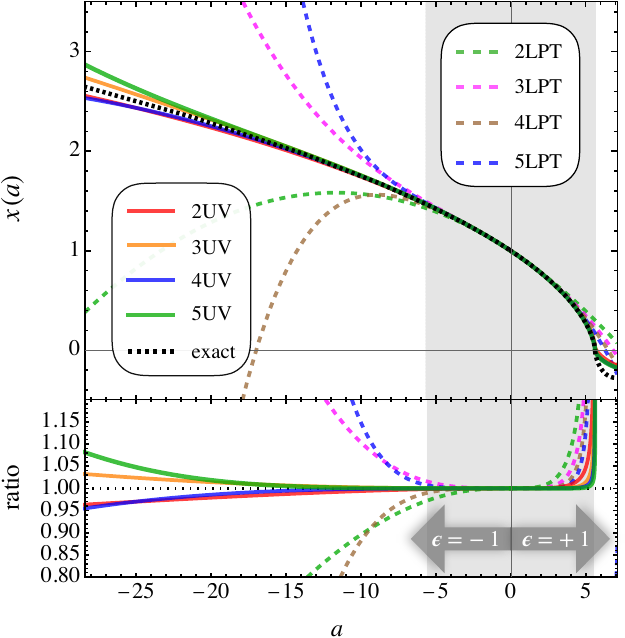}
   \caption{Same as Fig.\,\ref{fig:pade} but shown are UV-completed results at various truncation orders (solid lines; cf.\ Eq.\,\ref{eq:psicompl}) for the cases $\epsilon = \pm 1$ corresponding to $k = \pm 3/10$. For comparison, we have also added some $n$LPT results (dashed lines; cf.\ Eq.\,\ref{eq:x}).  
}   \label{fig:UV}
\end{figure}

In Fig.\,\ref{fig:UV} we show the resulting comoving matter trajectory for the so-obtained UV completion (solid lines) where, as before, the negative time branch corresponds to the void evolution. 
It is seen that the voids are most accurately described by the UV completion for the truncation order $n=3$ (orange line), whereas higher-order truncations become less accurate for large ``negative'' times. As mentioned above, this is an expected feature deep in the asymptotic regime. 
Still, any of the UV-completed predictions fare much better in comparison to LPT (dashed lines), at all times.
Furthermore, while the remainder of the LPT series (square bracketed term in~eq.\,\ref{eq:psicompl}) has the task of completing the LPT series to all orders, it clearly does not diverge in the void solutions.
This is made possible since the remainder is not limited by the use of a Taylor-series representation, as it is the case for LPT (or cosmological/Eulerian perturbation theory, for that matter).

Similar as with the RG methods, for the collapse case (positive time branch in Fig.\,\ref{fig:UV}) and within the convergent regime (gray shaded area), the UV completion performs increasingly better at increasingly higher-order truncations. We have explicitly verified this last statement by  determining the UV completions up to truncation orders of order one hundred (not shown).

Concluding this section, we have seen that the UV completed displacement most accurately resolves the spherical collapse and void evolution. This is made possible by exploiting the explicit knowledge of the LPT properties up to extremely high orders (see Fig.\,\ref{fig:domb}), which provided us with most accurate determinations of the two unknowns in the UV completion, namely the radius of convergence of the LPT series and the critical exponent appearing in Eq.\,\eqref{eq:psi-asymptotic}. Of course, such accurate knowledge of the unknowns cannot be expected ``in practice'', especially when the present UV method is extended to cosmological initial conditions. While a dedicated study to UV methods for cosmological initial conditions goes well beyond this initial study, we refer the interested reader to App.\,\ref{app:UV}, where we test UV predictions and expected errors when the two unknowns in the asymptotic method are less well estimated.

\section{Density evolution and distribution}\label{sec:density}

\subsection{Density evolution}\label{sec:NLdensity}

Given the various solutions in the asymptotic approaches, it is easy to evaluate their predictions of the corresponding nonlinear density contrast $\delta_\NL$, defined as
\be
  \delta_\NL (t) +1 =  |x_\NL(t)|^{-3} \,. 
\ee
Here, $x_\NL(t) = r_\NL(t)/a(t)$ is the comoving trajectory, where the subscript ``NL'' stands for a given nonlinear model prediction. 
Simple examples for this are
\begin{align}
 \deltaNL{1LPT} +1  &= \left( 1  - \delta_{\rm lin}/3 \right)^{-3} \,, \label{eq:deltaZA} \\
 \deltaNL{1RG\phantom{l}}  +1  &= \left( 1  - \delta_{\rm lin}/2 \right)^{-2} \,, \\
 \deltaNL{2RG\phantom{l}}  +1  &= \left( 1  - \delta_{\rm lin}/2 - 5 \delta_{\rm lin}^2/168 \right)^{-2} \,, \\
 \deltaNL{fit} + 1&=  \left( 1  - \delta_{\rm lin}/\alpha \right)^{-\alpha} \,, \label{eq:Bernie}
\end{align}
where $\delta_{\rm lin} := (3\epsilon/10) a$, and we note that $k = 3\epsilon/10$. Note that by construction, the leading-order Taylor expansion of the r.h.s.\ is always $1+\delta_{\rm lin}  + O(\delta_{\rm lin}^2)$. Equation~\eqref{eq:Bernie} traces back to work by Bernardeau~\cite{Bernardeau:1992,Bernardeau:1994,Bernardeau:1995} (it is a fit except in the limit $\Omega, \Lambda \to 0$), and the appearing parameter was originally set to $\alpha=3/2$.
Later on, it was found that the value of $\alpha=5/3$ produces a better fit (see e.g.\ \cite{Protogeros:1997,Mohayaee:2006,2008MNRAS.386..407L,Klypin:2018}). This formula has seen widespread use when relating linear and nonlinear densities, see e.g.\ Ref.\,\cite{Mohayaee:2006}, but particularly also to correct LPT displacements on small scales in hybrid methods, see e.g.\  Ref.\,\cite{Kitaura:2013}.

\begin{figure}
 \centering
   \includegraphics[width=1\columnwidth]{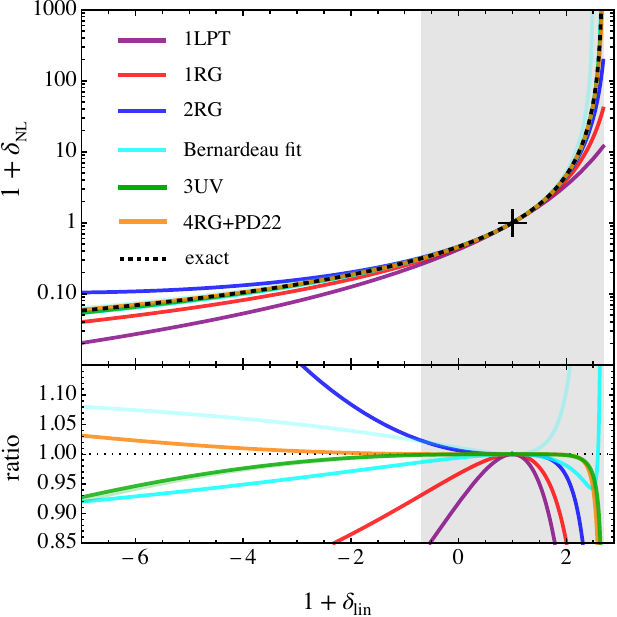}
   \caption{{\it Top panel:} Evolution of the nonlinear density contrast as a function of $1+\deltalin$.
  The asymptotic approaches 3UV (green line, solid with $\nu=2/3$ and faint with $\nu \simeq 0.675$) and 4RG+PD22 (orange) reproduce the exact result to good accuracy over a wide range of temporal scales. The cross marks the point of uniformity when $\deltaNL{NL} = 0 = \deltalin$, while the density singularity is reached when $\delta_{\rm lin} = \delta_{\rm c} \simeq 1.686$.
    The cyan solid [faint] line is based on the fit~\eqref{eq:Bernie} with $\alpha = 5/3$ [$\alpha = 3/2$].
    {\it Bottom panel:} Ratio of approximation versus the exact parametric solution.
}   \label{fig:density}
\end{figure}

For the two more elaborate asymptotic methods, the nonlinear densities read 
\begin{align}
  &\deltaNL{3UV\phantom{l}}  +1  = \bigg(  1- \frac{\deltalin}{3}  - \frac{\deltalin^2}{21}  + \frac{23\delta_{\rm c}}{756} \bigg\{  \deltalin^2 + 6 \deltalin \delta_{\rm c}   \nonumber \\
 &\qquad \hspace{1.1cm}+ 9 \left[  \left( 1 - \sdfrac{\deltalin}{\delta_{\rm c}} \right)^{2/3} \delta_{\rm c}^2 - 1 \right]  \bigg\}  \bigg)^{-3}, 
\intertext{where we set $\nu=2/3$ and $\delta_{\rm c} = (3/5) (3\pi/2)^{2/3} \simeq 1.686$, and}
  &\deltaNL{RG+PD22} + 1= 49 \left( 1553904 + \deltalin [10415 \deltalin -491508] \right)^2 \nonumber \\
  &\quad \times \left( 10877328 + \deltalin [ 1469453 \deltalin -8879220 ] \right)^{-2} .
\end{align}
In Fig.\,\ref{fig:density} we show a comparison of the resulting predictions for the nonlinear density. 
For the 3UV-completed result we show both the results for the critical exponent $\nu=2/3$ (solid green line) as well as for the approximative value $\nu \simeq 0.675$ (faint green line) as obtained from the Domb--Sykes extrapolation.
The former critical exponent leads to a slightly better density prediction in voids as well as in collapsing regions. Indeed, 
for $\nu=2/3$ [$\nu \simeq 0.675$] the linear-density prediction at collapse agrees with the exact result to $0.03\%$ [$0.058\%$] accuracy.
For our flagship prediction in the RG approach (4RG+PD22 shown in orange), the accuracy of the linear-density prediction at collapse is only $1.2\%$ and thus slightly worse as compared to the UV method. However, in void regions, the situation is different, and 4RG+PD22 delivers the most accurate density predictions considered in this article.

\subsection{One-point distribution of density} \label{sec:PDF}

Finally, we test our UV and RG methods by means of the one-point probability distribution function (PDF) of the matter density. For this it is useful to define the nonlinear overdensity
\be
  \varrho :=  \delta_\NL +1  \,,
\ee
where $\delta_\NL = \delta_\NL(\deltalin)$ is the nonlinear density contrast given in the previous section for various theoretical predictions. 
Provided that the initial density distribution is Gaussian and the validity of the spherical collapse model, the PDF is \cite{2001MNRAS.328..257S,Bernardeau2002,2008MNRAS.386..407L,Klypin:2018}
\begin{align}
  \varrho^2 p(\varrho) &=  \frac{1}{\sqrt{2\pi}} \exp\left( -\frac{\delta_{\rm lin}^2}{2\sigma^2}\right)
     \frac{\dd (\delta_{\rm lin}/\sigma)}{\dd \ln \varrho}  \,, \label{eq:PDF}
\end{align}
where $\sigma^2$ is the variance of the top-hat filtered linear density contrast. 
We normalize the probability $p$ such that $\int_0^\infty p(\rho) \dd \rho$ integrates to unity, but we note that this could be altered, e.g., to accommodate for populations of collapsed objects in high-density peaks. We also note that $\sigma$ depends on the chosen smoothing radius or, equivalently, on the overdensity (or mass) of the enclosing volume. For simplicity, we assume an initial power spectrum of $P(k) \propto k^n$ for which $\sigma = \sigma_v \varrho^{-(n+3)/6}$, where $\sigma_v$ is a constant amplitude. Finally, apart from Eq.\,\eqref{eq:PDF}, there are alternative theoretical models for the PDF, such as obtained from excursion set theory \cite{1998MNRAS.300.1057S} (see also~\cite{Klypin:2018} for a highly related study). We leave the comparisons of various PDF predictions against numerical simulations for future work.

\begin{figure}
 \centering
   \includegraphics[width=0.99\columnwidth]{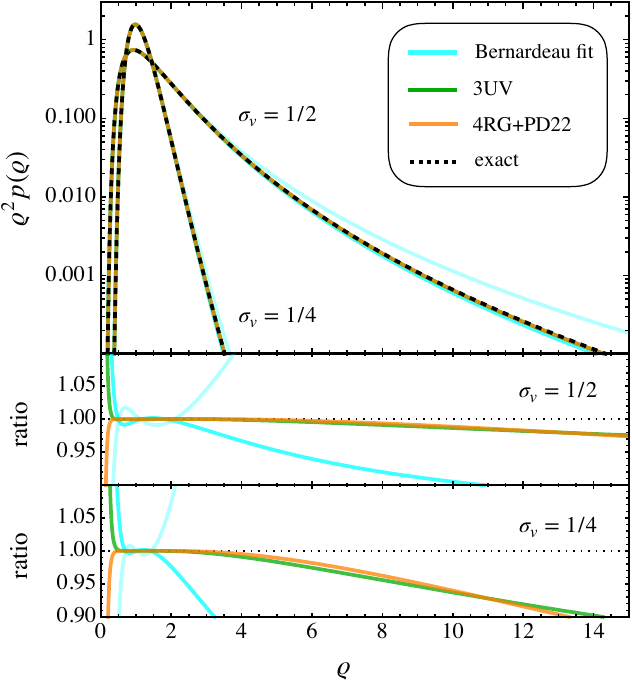}
   \caption{Same as Fig.\,\ref{fig:density} but shown is the normalized  density distribution function scaled with $\varrho^2$  for the filtering amplitudes $\sigma_v = 1/2,1/4$.
 We only show the best predictions to avoid cluttering.
}   \label{fig:PDF}
\end{figure}

In Fig.\ \ref{fig:PDF} we show the resulting PDF predictions  based on the RG and UV methods (we set $n\!=\!-2$ and $\sigma_v \!=\! 1/2,1/4$). Technically, this is achieved by inverting the functional relationship of $\varrho(\deltalin) = 1 + \delta_\NL(\deltalin)$ to $\deltalin(\varrho)$, such that the linear density contrast appearing in~\eqref{eq:PDF} is expressed in terms of the nonlinear overdensity.
From the figure it is seen that the 3UV (green line) and 4RG+PD22 (orange) predictions deliver nearly identical results for the PDF, and that they agree excellently with the exact prediction (black dashed line) over a wide range of overdensities.

\section{Summary and conclusions}\label{sec:concl}

\noindent {\bf Technical summary.}
Lagrangian perturbation theory is known to converge fast for gravitational collapse that predominantly occurs along one coordinate axis \cite{2017MNRAS.471..671R,2018PhRvL.121x1302S,2022A&A...664A...3S}---which is a generic feature of triaxial collapse, giving rise to Zel'dovich's pancakes.
This situation changes to the opposite when considering the case of spherically symmetric collapse \cite{1996MNRAS.282..641S,1997A&A...326..873K,2007PhRvD..75d4028T,2019MNRAS.484.5223R}, for which LPT convergence is very slow.
Even worse, the LPT series for the void evolution displays divergent behavior after a critical time (see e.g.\ Fig.\,\ref{fig:rainbow}), although the trajectories should stay perfectly smooth.
 This pathological behavior suggests that the Taylor-series approach inherent in LPT is vastly inefficient for resolving spherical collapse.

In the present article, we have analyzed two asymptotic approaches that remedy these drawbacks of LPT.
One of the avenues employs renormalization-group techniques, where  the evolution equation is first recast such as to identify an effective expansion parameter---in the present case the rescaled spatial curvature parameter $\epsilon =10k/3$. The evolution equation is then solved with a na\^ive perturbation {\it Ansatz} in powers of $\epsilon$. Subsequently, 
the perturbation equations are renormalized in order to expose the secular terms, i.e., terms that would grow unboundedly for large times. After evaluating the RG-flow condition, which removes the arbitrariness of certain integration constants, one then obtains the final renormalized result. 
These steps have been effectively implemented in a particularly concise way in Sec.\,\ref{sec:RG}; we refer the interested reader to App.\,\ref{app:RG+} for more elaborate technical details.

The analyzed RG technique predicts already in the first iteration a singular behavior of the solution with the correct critical exponent within the leading-order asymptotics (cf.\ App.\,\ref{app:para} for complementary asymptotic considerations by means of the exact parametric result). Subsequent higher-order iterations within the RG method are particularly fruitful when paired with Pad\'e approximants (Sec.\,\ref{sec:HotRG-Pade}). 
Overall, we find that the Pad\'e approximant (2,2) of the 4RG prediction (dubbed 4RG+PD22) delivers the most accurate RG result. We find some evidence that Pad\'e approximants are not just better approximations, but in fact are crucial to restore a symmetry property of the solution which, in the present case, would otherwise lead to unphysical collapse of void solutions.

The other asymptotic approach that we considered is the UV completion (Sec.\,\ref{sec:asy-UV}). For this we first analyzed the large-order asymptotic properties of the LPT series, for which we drew the so-called Domb--Sykes plot (Fig.\,\ref{fig:domb}). From this plot it is seen that subsequent ratios of LPT coefficients settle into a linear relationship at large orders, from which one can deduce that the radius of convergence of the LPT series is limited by a singularity at the (real-valued) collapse time. Even more, the graphical analysis reveals a measured critical exponent that closely resembles the one obtained from the RG method.
With this input at hand, one can complete the LPT series in a highly efficient manner (eq.\,\ref{eq:psicompl}). Crucially, the UV-completed prediction does not come with the above mentioned pitfalls of LPT, since the remainder of the series is not a Taylor series anymore.
We note however that the UV completion should be performed at sufficiently low truncation orders, 
otherwise unwanted features in the void evolution are ``reactivated''.
For both over- and underdense regions, we find that the UV completion with third-order truncation in LPT (dubbed 3UV) leads to the most convincing predictions.

Finally, we have tested our two asymptotic methods by means of predicting the nonlinear density evolution and corresponding one-point distribution (Sec.\,\ref{sec:density}).
Characteristically, we find that the UV method works marginally better for predicting the nonlinear density near the collapse (which comprises a huge challenge for LPT) in comparison to our flagship RG method, while the situation is opposite in voids; see Fig.\,\ref{fig:density}.
In contrast, when predicting the  one-point probability distribution function of matter, we found only negligible differences between the tested UV and RG methods (Fig.\,\ref{fig:PDF}). \\[0.3cm]

\noindent {\bf Concluding remarks.} 
We have analyzed the dynamical process from gravitational infall to collapsed structures from the perspective of a classical phase transition. We have seen clear indications
that fixed-order LPT is ultimately limited in capturing related critical phenomena.
For the idealized case of spherical collapse, we provide two ways
 to remedy the situation. 
The crucial step in both cases is to incorporate a non-analytic term $\propto (\mathfrak{a}_\star-\mathfrak{a})^\nu$ that captures the critical nature of gravitational collapse, where $\nu$ is the critical exponent, and the (curvature rescaled) cosmic scale factor $\mathfrak{a}= a k$ takes the role of the order parameter.

The obvious challenge of the discussed asymptotic methods is their potential applicability for predicting the evolution of collisionless matter for cosmological initial conditions in three space dimensions.
For the UV method, one first needs to find the precise nature of the convergence limiting singularity(ies) for the Lagrangian displacement field, which requires high-order LPT solutions that are however already available \cite{2014JFM...749..404Z,2015PhRvD..92b3534M,2021JCAP...04..033S}. In this context, we note that in Ref.\,\cite{2021MNRAS.501L..71R}, we obtained some numerical evidence that the norm of the displacement  
contains a non-analytic term that is in structure formally identical to the one in the case of spherical collapse.

Regarding a UV completion within an Eulerian-coordinates formulation,  
it is likely that the leading-order asymptotic behavior of the fluid variables is characterized by a pair of complex-conjugated singularities in time; see Ref.\,\cite{2022arXiv220712416R} for highly related avenues applied to the inviscid Burgers equation. 
However, in such a case, the asymptotic structure of the fluid variables would be just $\propto  (a_\star - a)^\nu + (\bar a_\star - a)^\nu$, where $a_\star$ is now complex and $\bar a_\star$ denotes its complex conjugate. Hence, the remainder of the respective series has a different structure as in the spherical case, but this can easily be handled by suitable alterations within the framework. 

Similarly, the RG method needs to be suitably adapted when applied to cosmological initial conditions. First of all, the underlying fluid equations are partial differential equations in time {\it and} space. Therefore, as a preparatory step, the fluid equations could be first represented in a Fourier basis, which then leads to a spatially decoupled ODE in time for each Fourier mode of the fluid variable. The solutions to these ODEs could then be renormalized by the methods outlined in this paper, {\it provided a suitable expansion parameter is identified}. Regarding the latter, we remind the reader that our starting point for the RG method, Eq.\,\eqref{eq:RGmain},  is obtained by time integrating the fluid equations for spherical collapse. This time integration then revealed as an integration constant the spatial curvature, which indeed acts as  the expansion parameter in the present RG approach. Whether similar derivations hold for the Fourier coefficients of the fluid variables for random initial conditions remains to be investigated.

Future applications of the asymptotic methods include the accurate modeling of (non-spherical) void and overdense regions in deterministic or statistical contexts (e.g.\ excursion sets and data inference for  field-level forward modelling), as well as for hybrid approaches where the methods are paired with numerical simulation- or machine learning techniques. 
Lastly, in this paper we did not consider post-shell-crossing effects \cite{2017MNRAS.470.4858T,2018JCAP...06..028P,2020JCAP...06..033C,2021MNRAS.505L..90R,2022A&A...664A...3S},  but the RG and UV methods are in principle able to handle such critical behavior, which however requires further investigation.
In the long term, asymptotic methods have the potential for reducing the gap between theoretical and numerical methods, while at the same time enhance the physical insight into highly nonlinear problems.

\begin{acknowledgments}
We thank Hamed Barzegar and Sharvari Nadkarni-Ghosh for useful discussions. 
\mbox{O.H.~acknowledges funding from the} European Research Council (ERC) under the European Union’s Horizon 2020 research and innovation program, Grant Agreement No. 679145 (COSMO-SIMS).
\end{acknowledgments}

\appendix

\section{Exact parametric solution} \label{app:para}

For completeness we report here the exact parametric solution for the case of spherically symmetric over- and underdensities. Most of the details are well known, see e.g.~\cite{1967ApJ...147..859P,1969PThPh..42....9T,1972ApJ...176....1G,1994ApJ...431..486B}), but we also provide an asymptotic analysis by means of the parametric result that is perhaps less well known. 

Consider the parametrized evolution of a spherical density perturbation in an expanding universe with vanishing cosmological constant. By Birkhoff's theorem, the interior perturbation can be described in isolation as a separate universe of constant scalar curvature $K=10k/3$ obeying 
\be
 \left( \frac{\dot{R}}{R} \right)^2 = \frac{2M}{R^3} - \frac{K}{R^2}  \,.
\ee
Let us change to conformal time $\eta$ defined via $\dd \eta  =  \dd t/R$ and non-dimensionalize then with $\hat{R}:=M$, so that we have the new dimensionless radius coordinate $r := R/\hat{R}$ of the form
\be
 \left( \frac{{\rm d}r}{{\rm d}\eta}\right)^2 = 2r - Kr^2 \,.
\ee
Given the boundary condition at $r(0)=0$, 
the general solution in conformal time is $ r(\eta) =  [1- \cos( \sqrt{K} \eta)]/K$, and for the specific (limiting) cases
\be \label{eq:reta}
 r(\eta) = 
 \left\{ 
  \begin{array}{lll}
   1-\cos \eta & \textrm{for} & K=+1 \\
   \eta^2/2 & \textrm{for} & K=0 \\
   -1+\cosh \eta & \textrm{for} & K=-1
  \end{array}
 \right. \,.
\ee 
Instead of having solutions in terms of conformal time, we like to express these solutions in terms of either cosmic time~$t$ or in terms of the scale factor $a$ of a flat background cosmology. 
Integrating the relation   ${\rm d}t = R(\eta){\rm d}\eta$ yields
$t(\eta) = \eta/ K - \sin(\sqrt{K} \eta)K^{-3/2}$,
 and for the specific cases
\begin{align}
t(\eta) &=  
  \left\{
    \begin{array}{lll}
      \eta-\sin \eta , & \, K=+1 \\
      \frac{\eta^3}{6} , & \, K=0 \\
      -\eta+\sinh \eta , & \,  K=-1 
    \end{array}
  \right. \,. \label{eq:teta}
\end{align}
In a spatially flat, matter dominated universe we can write
\begin{align}
a(t) = \frac{1}{2} (6t)^{2/3} \,.
\end{align}
Using this relation combined with the above solution for $t(\eta)$, we then have
\begin{align}
 a(\eta) &= 
  \left\{ 
    \begin{array}{lll}
      (1/2)\left[ 6(\eta-\sin \eta)\right]^{2/3} , &  \, K=+1 \\
      (1/2)\eta^2 ,  & \, K=0 \\
      (1/2)\left[ 6(\sinh \eta-\eta)\right]^{2/3},  &  \, K=-1
    \end{array}
   \right. \,. \label{eq:aeta}
\end{align}
On a technical level, to get the solution $r$ as a function of~$a$, we plot parametrically $r(\eta)$ against $a(\eta)$. 
In the positive curvature case, the first shell-crossing occurs for $\eta=2\pi$, which translates to the known result 
\be
  a_\star = (3\pi\sqrt{2})^{2/3} \simeq 5.622 \,.
\ee
Furthermore, we derive the critical exponent of the shell-crossing singularity by means of the above-mentioned parametric solution; see Refs.\,\cite{Penston:1969,Nakamura:1985,Gurevich:1995,White:2022} for highly related avenues.

For this we consider the Taylor expansion of
\be \label{eq:rofetaof}
  r(\eta (a)) = 1 - \cos \eta(a)  
\ee
around the shell-crossing time $a(\eta)|_{\eta = 2\pi} =a_\star$.
Here,  $\eta(a)$ is the inverse function of $a(\eta) =  (1/2)[ 6(\eta-\sin \eta)]^{2/3}$ which is  {\it a priori} unknown in explicit form, however for evaluating the Taylor expansion of~\eqref{eq:rofetaof} around the shell-crossing time, only the series reversion is needed. One straightforwardly finds
\be \label{eq:AsymSC}
  r(\eta(a))|_{a = a_\star} = c \left(a-a_\star \right)^{2/3}  + O\left((a-a_\star)^{4/3}  \right) \,,
\ee
with $c = 3^{8/9} \pi^{2/9} 2^{-5/9}$, thereby identifying a critical exponent of $2/3$ in the leading-order asymptotics. Quite astonishingly, already the first-order renormalization-group approach from section~\ref{sec:RG} correctly predicts this critical exponent.

\begin{figure}
 \centering
   \includegraphics[width=0.99\columnwidth]{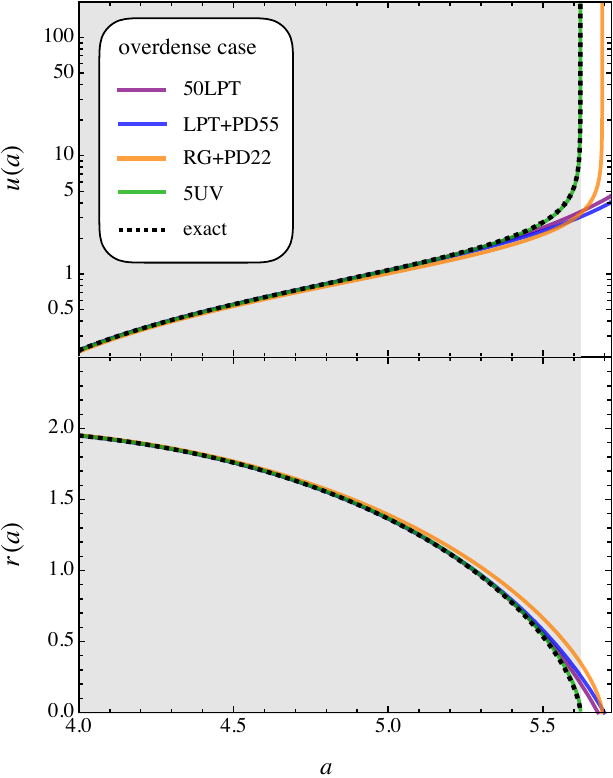}
   \caption{{\it Top panel:} the velocity $u := - \partial_a r$ in the overdense case for $\epsilon = 1$ and $k=3/10$. 
Note that we do not show the full temporal evolution but focus on times close to the blowup, which in the present case is at $a = a_\star\simeq 5.622$. 
The depicted RG + Pad\'e approach (orange line; based on Eq.\,\ref{eq:PD22}) as well as the UV method (green line) reproduce the generic singular behavior as expected from the exact spherical collapse model (black dashed line). By contrast, 50LPT (purple line) or its plain Pad\'e  approximant ``5,5'' (blue line) do not predict this behavior. 
   {\it Bottom panel:} same but showing the physical trajectory for comparison. 
  } \label{fig:appVR}
\end{figure}

Finally, for completeness we show in the top panel of Fig.\,\ref{fig:appVR} the velocity $u := - \partial_a r$ (minus sign is convention) for the exact prediction (black dashed lines), as well as for several approximation schemes (various colors). Clearly, the exact parametric solution, the RG- and UV approaches detect the emergence of a singular velocity at collapse time $a=a_\star$ (albeit with an unwanted shift of the RG result along the temporal axis that could be refined at higher orders). As mentioned earlier, this singular behavior is expected, simply due to the asymptotic nature of the problem. Still, we should point out that it is in essence the chosen temporal variable (here the scale factor) that causes this singularity. Indeed, taking the conformal time as the temporal parameter for the physical trajectory, it is trivially seen that the corresponding velocity $\hat u =- \partial_\eta r(\eta)$ remains finite at all times in the overdense case (cf.\ Eq.\,\ref{eq:reta}).
Thus, the temporal coordinate transformation~\eqref{eq:teta} acts as the de-singularization transformation for spherical collapse.
 Hence, much in the sense as for the singularity of the Schwarzschild solution at the horizon of a black hole, also the present singularity is removable and thus not really physical. 
Note however that new singularities might be easily introduced, once the present analysis is extended into the post-collapse regime; see~\cite{2017MNRAS.470.4858T,2021MNRAS.505L..90R,2022A&A...664A...3S} for related avenues however within a one-dimensional setup.

\section{More details to the RG approach} \label{app:RG+}

Here we reconsider the RG calculations from Sec.\,\ref{sec:RG} but follow more closely the formalisms of Refs.\,\cite{1994PhRvL..73.1311C,Chen:1996,2008PhyD..237.1029D}, who applied  RG techniques to various non-cosmological fluids.
On top of that, we will apply suitable multiscaling techniques that in fact allows us to solve the corresponding RG flow equations {\it exactly}. Nevertheless, we should stress that the reported results agree exactly with those reported in Sec.\,\ref{sec:RG}. Thus, the following demonstrations  serve rather as material for the reader to gain more intuition about the applied asymptotic methods.

As stated in the main text, solving the evolution equation
\be  \label{eq:RGmainREP}
  r'^2 = \frac a r - \epsilon \frac a 2
\ee
 by the perturbative {\it Ansatz} $r = r_0 + \epsilon \, r_1 + \ldots$, one obtains at the zeroth order
\begin{align}
  &\epsilon^0 \bigg\{ r_0'^2 = \frac{a}{r_0}  \bigg\}  \label{eq:evoRG0rep} \,.
\end{align}
In Sec.\,\ref{sec:RG} we solved this ODE without explicit boundary conditions; its solution can be written as
\be \label{eq:r0repAPP}
   r_0 = \left(  c_2 + a^{3/2} \right)^{2/3},
\ee
where $c_2$ is an integration constant. From here on we will execute two main alterations as compared to the approach of Sec.\,\ref{sec:RG}. The first alteration performs a multiscale zoom that is explained in the following section.
The second alteration is related to an alternative RG approach that takes explicit boundary conditions for the ODE into account.

\subsection{Multiscaling refinement}

Observe that Eq.\,\eqref{eq:r0repAPP} contains two exponents: the interior exponent is $3/2$ while the exterior exponent is $2/3$. These exponents suggest that the evolution equation~\eqref{eq:RGmainREP} can be written in a more efficient way, provided that we employ the rescaled spatiotemporal coordinates
\be \label{eq:scaling}
  R^{2/3} := r \,, \qquad T := a^{3/2} \,.
\ee
With these changes, Eq.\,\eqref{eq:RGmainREP} becomes
\be
  \boxed{ \dot R^2 = 1 - \frac{\epsilon}{2} R^{2/3}  }\,, \label{eq:zoomedODE}
\ee
where a dot denotes a time derivative w.r.t.\ (cosmic) time $T$.
Equation~\eqref{eq:zoomedODE} is autonomous in the time variable, which is clearly not the case for its parent equation~\eqref{eq:RGmainREP}. Furthermore, the strong nonlinearity of $1/r$ in~\eqref{eq:RGmainREP} has been converted into a weaker nonlinearity~$\sim R^{2/3}$ in~\eqref{eq:zoomedODE}. These combined observations suggest that perturbative techniques in the scaled spatiotemporal coordinates may grasp already the dominant asymptotic features for the given problem.

Therefore, in  what follows we seek perturbative solutions for~\eqref{eq:zoomedODE}, as opposed to solutions to its parent equation~\eqref{eq:RGmainREP}.
We remark that the outlined RG method works also without employing the rescaled coordinates, however the corresponding RG flow equation can then only be solved perturbatively.

To solve~\eqref{eq:zoomedODE} we impose a na\^ive perturbation {\it Ansatz} for the rescaled trajectory
\be
  R = R_0 + \epsilon R_1 + \epsilon^2 R_2 + \ldots \,,
\ee
which leads to the following perturbation equations
\begin{subequations} \label{eqs:pertEqsrescaled}
\begin{align}
  &\epsilon^0 \Bigg\{  \dot R_0^2 = 1  \Bigg\} \,, \label{eq:eps0Scaling}  \\
 &\epsilon^1 \Bigg\{  2 \dot R_0 \dot R_1 = - \tfrac 1 2 R_0^{2/3}  \Bigg\} \,,  \\
 &\epsilon^2 \Bigg\{  \dot R_1^2 + 2 \dot R_0 \dot R_2 = - \frac{R_1}{R_0^{1/3}}  \Bigg\} \,,
\end{align}
\end{subequations}
etc. Now, if one solves these equations with the boundary conditions $R(0)=0$, one immediately  
obtains 
\be \label{eq:solScaled}
  R = T - \frac{3\epsilon}{20} T^{5/3} - \frac{3\epsilon^2}{1120} T^{7/3} + O(\epsilon^3)\,,
\ee
which is equivalent in the unscaled coordinates with
\be
  r= a \left(  1- \frac{3\epsilon a}{20} - \frac{3 \epsilon^2 a^2}{1120} \right)^{2/3}  + O(\epsilon^3) \,.
\ee
This result agrees exactly with~\eqref{eq:RG2sol}.
At this point we could stop the derivation as the result coincides already with the one stated in the main text. Instead we continue even further, however, now with the actual RG procedure. We will find that the RG procedure on top of the multiscaling approach leads to no further improvement,
indicating that the RG techniques used cannot be further improved for the present task (this is why we apply Pad\'e approximants to the RG approach).

\subsection{Refined RG method}\label{app:refinedRG}

Here we apply an RG method that is in the spirit of the seminal approaches of Refs.\,\cite{1994PhRvL..73.1311C,Chen:1996,2008PhyD..237.1029D}. In contrast to the simplified approach outlined in Sec.\,\ref{sec:RG}, here we enable explicit boundary conditions to 
solve the perturbation equations~\eqref{eqs:pertEqsrescaled}, provided at an arbitrary time $T= T_0$. The appearing integration constants in the solutions are then renormalized with the aim to isolate the term(s) that grow unboundedly for large times. But $T_0$ is in an arbitrary timescale and the solution should not depend on it. Hence one demands an RG flow condition that in essence removes this arbitrariness from the solution.

Another interpretation of the RG flow condition is as follows \cite{2008PhyD..237.1029D}. Suppose the solution $R(T)$ has been obtained by demanding the initial condition ${\cal R}(T_0)$ at arbitrary initial time~$T_0$. The solution is thus $R = R(T, T_0, {\cal R}(T_0))$ where the implicit dependence of $R$ on the initial data can be understood as {\it characteristics along the solution curve.} 
Put differently, the solution $R(T, T_0, {\cal R}(T_0))$ is identical to the solution $R(T, T_1, {\cal R}(T_1))$ with $T_1 \neq T_0$, as long as both initial values  ${\cal R}(T_0)$ and  ${\cal R}(T_0)$ lie on the same solution curve of~$R$.
It can then be shown that the RG condition  $\dd R/ \dd T_0 = 0$ leads exactly to the parent ODE, however now not formulated for $R$ but directly for the characteristics~${\cal R}(T_0)$.

Let us begin with the calculational steps.
Solving Eq.\,\eqref{eq:eps0Scaling} with the initial condition $R_0(T_0) = {\cal R}_0$ we obtain
\begin{align}
  R_0 &= T - T_0 + {\cal R}_0 \,.
\intertext{The next-order perturbative equations are solved with the initial conditions $R_1(T_0)=0 = R_2(T_0)$. We get}
  R_1 &= -\frac{3}{20} \left[  (T- T_0 + {\cal R}_0)^{5/3} - {\cal R}_0^{5/3} \right] \,, \\ 
  R_2 &= - \frac{3}{1120} \Big[  (T- T_0 + {\cal R}_0)^{7/3}  \nonumber \\
   & \qquad + 14 {\cal R}_0^{5/3} (T- T_0 + {\cal R}_0)^{2/3}  - 15 {\cal R}_0^{7/3}  \Big] \,.
\end{align}

\noindent{\bf First-order renormalization.} We first focus on the solution for $R$ valid to order $\epsilon$, which
reads
\be
  R = T - T_0 + {\cal R}_0  - \frac{3\epsilon}{20} \left[  (T- T_0 + {\cal R}_0)^{5/3} - {\cal R}_0^{5/3} \right] .
\ee
Here the secular (i.e., strongest growing term in time) is located within the square bracket. To isolate the secular term, we employ a renormalized integration constant ${\cal R}(T_0)$ that absorbs the nonsecular term in the square bracket. This is achieved by the transformation 
\be \label{eq:Rren1}
 {\cal R}_0 = {\cal R}(T_0) - \frac{3\epsilon}{20} {\cal R}^{5/3}(T_0) \,.
\ee
Employing ${\cal R}(T_0)$, the renormalized solution for $R$ becomes
\be \label{eq:renScaled}
  R = T - T_0 + {\cal R}(T_0)  - \frac{3\epsilon}{20}   \left[T- T_0 + {\cal R}(T_0) \right]^{5/3}  \,,
\ee
to $O(\epsilon^2)$. Since the arbitrary timescale $T_0$ does not appear in the original problem, we impose 
the RG flow equation
\be
  \left. \frac{\dd R}{\dd T_0}  \right|_{T_0 = T} =0 \,,
\ee
which leads exactly to (i.e., no expansion needed!)
\be
  {\cal R}(T) = T + C_1 \,,
\ee
where $C_1$ is an arbitrary integration constant that, in the present case, just shifts the temporal coordinate. Setting the shift $C_1$ to zero, one finds the renormalized solution
\be
  R = T - \frac{3 \epsilon}{20} T^{5/3} \,,
\ee
which, after reverting the scaling~\eqref{eq:scaling}, agrees with the first-order renormalized result~\eqref{eq:1RGsol} in the main text.

We remark that the above RG procedure could be alternatively executed by introducing a new time $\tau$ and write $T-T_0 = T- \tau + \tau - T_0$ in~\eqref{eq:renScaled}. Subsequently, one absorbs the terms proportional to $\tau - T_0$ into a new renormalized integration constant, and evaluates the altered RG flow equation  $\dd R/\dd \tau |_{\tau =T}=0$. This procedure, which would be closer in the original spirit of Refs.\,\cite{1994PhRvL..73.1311C,Chen:1996},  leads however to equivalent results as stated above.

\vspace{0.3cm}
\noindent{\bf Second-order renormalization.}
Similarly steps as above can be executed at higher orders. We begin with
the (fully) un-renormalized  solution 
\begin{align} \label{eq:un-ren2}
  R &= T - T_0 + {\cal R}_0  - \frac{3\epsilon}{20} \left[  (T- T_0 + {\cal R}_0)^{5/3} - {\cal R}_0^{5/3} \right] \nonumber \\
 &\quad    - \frac{3\epsilon^2}{1120} \Bigg[  (T- T_0 + {\cal R}_0)^{7/3} - 15 {\cal R}_0^{7/3}  \nonumber \\
 &\qquad\hspace{1.5cm} + 14 {\cal R}_0^{5/3} (T- T_0 + {\cal R}_0)^{2/3}   \Bigg]
\end{align}
up to order $\epsilon^2$.
From the considerations of the first-order renormalization, we know already the transformation of ${\cal R}_0$ to order $\epsilon$; see Eq.\,\eqref{eq:Rren1}. To make progress we use this result and set ${\cal R}_0 = {\cal R}(T_0) - (3\epsilon/20) {\cal R}^{5/3}(T_0) + \epsilon^2 a_2$, where $a_2$ is an unknown. Plugging this relationship for ${\cal R}_0$ into~\eqref{eq:un-ren2} leads to the perturbation equations
\begin{align}
 R_0 &= T -T_0 + {\cal R} \,, \\
 R_1 &= -\frac{3}{20} (T- T_0 + {\cal R})^{5/3} \,, \\
 R_2 &= a_2  - \frac{3}{1120} \left[ (T- T_0 + {\cal R})^{7/3}  -  {\cal R}^{7/3} \right] \,, \label{eq:R2}
\end{align}
where  ${\cal R}(T_0) := {\cal R}$ for brevity.
Notice that the second-order term~$\sim (T-T_0 + {\cal R}_0)^{2/3}$ in~\eqref{eq:un-ren2} has disappeared thanks to the first-order renormalization.
To complete the second-order renormalization, we set $a_2 = - 3 {\cal R}^{7/3}/1120$, which removes in~\eqref{eq:R2} the term that purely depends on the initial condition, thereby isolating the remaining secular term at order $\epsilon^2$. In summary, we can thus write Eq.\,\eqref{eq:un-ren2}  as
\begin{align}
  R &=  T -T_0 +  {\cal R}  -\frac{3\epsilon}{20} (T- T_0 + {\cal R})^{5/3}  \nonumber \\
  &\qquad   - \frac{3\epsilon^2}{1120}  (T- T_0 + {\cal R})^{7/3} + O(\epsilon^3) \,.
\end{align}
Imposing the RG flow equation $\dd R/\dd T_0 |_{T_0 = T} =0$ leads to the exact solution  ${\cal R}(T) = T + C$ as obtained from the first-order renormalization, indicating that the renormalization procedure is consistent. In summary,
the renormalized solution reads
\be
  R = T - \frac{3\epsilon}{20} T^{5/3} - \frac{3\epsilon^2}{1120} T^{7/3}  + O(\epsilon^3)
\ee
or, in unscaled coordinates,
\be
  r= a \left(  1- \frac{3\epsilon a}{20} - \frac{3 \epsilon^2 a^2}{1120} \right)^{2/3}  + O(\epsilon^3) \,,
\ee
which agrees with the 2RG result stated in the main text.

\begin{figure}[!t]
 \centering
   \includegraphics[width=0.9\columnwidth]{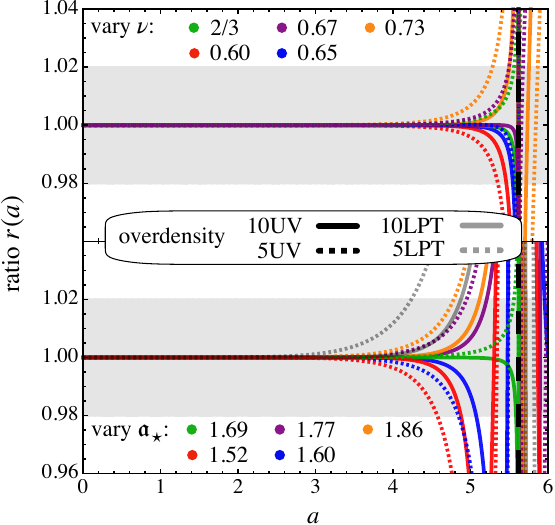}
   \caption{Evolution of physical trajectory $r(a)$ in the collapse case ($k=3/10$). For convenience we show the ratio of the various method versus the exact parametric result, and the vertical long-dashed black line denotes the time of collapse. {\it Top panel:} the critical exponent is varied while the radius of convergence~$\mathfrak{a}_\star$ is fixed to the correct value.
{\it Bottom panel:} the critical exponent is fixed to $\nu=2/3$ and the value of $\mathfrak{a}_\star$ is varied.  We also show the 10LPT [5LPT] predictions for comparison.
}   \label{fig:ratiovaryover}
\end{figure}

\section{Further results to UV completion} \label{app:UV}

Here we provide further results within the framework of UV completion. Specifically, 
we analyze the predictions for the case when the two unknowns in the method are not known to high precision, which is particularly relevant when the UV completion is applied to random field initial conditions.
As a reminder, these two unknowns are the critical exponent~$\nu$ and the radius of convergence of the LPT series $\mathfrak{a}_\star$, which appear within the present UV completion as follows,
\be \label{eq:PsiAsyApp}
   \psiASY(a) = A  \left( \mathfrak{a}_\star - \mathfrak{a} \right)^\nu \,, \qquad \mathfrak{a} = a k 
\ee
(the constant $A$ is fixed by the UV matching criterion).
In Fig.\,\ref{fig:ratiovaryover} we show the temporal evolution of the physical trajectory in the overdense case. In the top [bottom] panel we fix $\mathfrak{a}_\star= (3/5) (3\pi/2)^{2/3} \simeq 1.686$ [$\nu =2/3$]  while we vary the critical exponent [radius of convergence]. Clearly, varying these parameters affects the prediction of the final stages of the collapse (at $a\simeq 5.622$, vertical black dashed line) significantly, while at earlier times, the impact is very small.

Generally, the prediction of the overdense collapse is more hampered by a potential lack of precise knowledge on $\mathfrak{a}_\star$ than on $\nu$. Still, even if $\mathfrak{a}_\star$ is underpredicted by more than 10\%, the UV prediction fares still better than the respective LPT prediction at the same perturbation order.

\begin{figure}[!h]
 \centering
   \includegraphics[width=0.9\columnwidth]{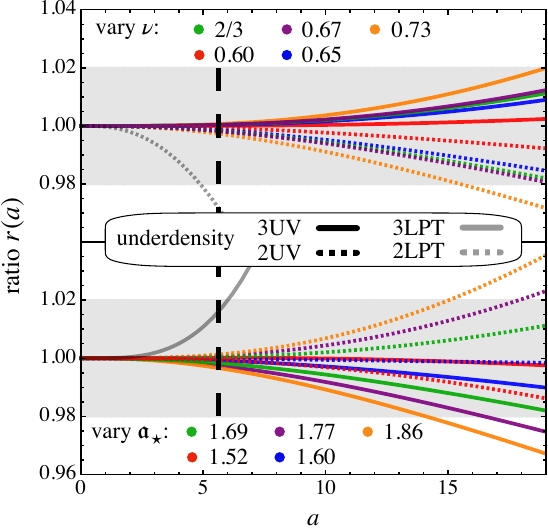}
   \caption{Same as Fig.\,\ref{fig:ratiovaryover} but showing the physical trajectory in the void case ($k=-3/10$).
}   \label{fig:ratiovaryunder}
\end{figure}

In Fig.\,\ref{fig:ratiovaryunder} we show the same as above but now applied to the void case. Here, 
the UV completion shows a fairly weak dependence on both $\mathfrak{a}_\star$ and $\nu$; in fact, the mere knowledge of the structural form of the asymptotic form~\eqref{eq:PsiAsyApp} appears to be enough to clearly outperform the respective LPT predictions.

\bibliography{biblio.bib}

\end{document}